\documentclass{ws-p8-50x6-00}

 \catcode`@=11
\def\@cite#1#2{\unskip\nobreak\relax
    \def\@tempa{$\m@th^{\hbox{\the\scriptfont0 #1}}$}%
    \futurelet\@tempc\@citexx}
\def\@citexx{\ifx.\@tempc\let\@tempd=\@citepunct\else
    \ifx,\@tempc\let\@tempd=\@citepunct\else
    \ifx;\@tempc\let\@tempd=\@citepunct\else
    \let\@tempd=\@tempa\fi\fi\fi\@tempd}
\def\@citepunct{\@tempc\edef\@sf{\spacefactor=\the\spacefactor\relax}\@tempa
    \@sf\@gobble}
\catcode`@=12

\makeatletter
\def\citenum#1{{\def\@cite##1##2{##1}\cite{#1}}}
\def\title#1{\gdef\@title{#1}}
\makeatother
\let\rightnote\relax
\let\fbox\relax

\begin{document}

\font\fortssbx=cmssbx10 scaled \magstep1

\title{\vspace*{-.5in}
\hbox to \hsize{
\hbox{\fortssbx University of Wisconsin - Madison}
\hfill$\vtop{\normalsize\hbox{\bf MADPH-00-1168}
                \hbox{\rm March 2000}
                \hbox{\hfil}}$ }
\uppercase{Overview of Neutrino Oscillation Physics}\footnote{Talk presented at the {\it 7th International Symposium on Particles, Strings and Cosmology (PASCOS\thinspace99)}, Granlibakken, California, December 1999}}

\author{V. Barger}

\address{Physics Department, University of Wisconsin, Madison, WI 53706}

\maketitle

\abstracts{
Recent evidence for neutrino oscillations has revolutionized the study of neutrino masses and mixing. This report gives an overview of what we are learning from the neutrino oscillation experiments, the prospects for the near term, and the bright future of neutrino mass studies.
}

\section{Neutrino Masses}

Tree-level mass generation occurs through the Higgs mechanism. The Dirac mass $m_D$ arises  in a lepton conserving ($\Delta L = 0$) interaction  and requires a right-handed neutrino. A Majorana mass $m_M$ occurs through a $\Delta L = 2$ process with only a left-handed light neutrino field and a heavy isosinglet intermediate field $N^c$. Then the see-saw mechanism with $m_D \sim 10^2$~GeV and $m_M > 10^{12}$~GeV generates light neutrinos
\begin{equation}
m_\nu = {m_D^2/ m_M}
\end{equation}
that are nearly Majorana.
In the case that $m_D\sim m_M\sim\rm eV$, as can be realized in some models, active--sterile neutrino oscillations can take place.

Neutrino mass can alternatively be generated radiatively by new interactions, such as the $R$-parity violating $\nu b\tilde b$ interaction. Recent reviews of theoretical models for neutrino mass generation are given in Ref.~\citenum{RMreview}.

\section{Three-Neutrino Oscillations in Vacuum and Matter}

The relation of the three-neutrino flavor eigenstates to the mass eigenstates~is
\begin{equation}
\left(\begin{array}{c} \nu_e\\ \nu_\mu\\ \mu_\tau\end{array}\right) = 
U \left(\begin{array}{c} \nu_1\\ \nu_2\\ \nu_3\end{array}\right)\,,
\end{equation}
where $U$ is the $3\times3$ Maki-Nakagawa-Sakata (MNS) mixing matrix. It can be parametrized by
\begin{eqnarray}  \addtolength{\arraycolsep}{-.2em}
U {=}\!\! \left(\begin{array}{ccc}
  c_{13} c_{12}       & c_{13} s_{12}  & s_{13} e^{-i\delta} \\
- c_{23} s_{12} - s_{13} s_{23} c_{12} e^{i\delta}
& c_{23} c_{12} - s_{13} s_{23} s_{12} e^{i\delta}
& c_{13} s_{23} \\
    s_{23} s_{12} - s_{13} c_{23} c_{12} e^{i\delta}
& - s_{23} c_{12} - s_{13} c_{23} s_{12} e^{i\delta}
& c_{13} c_{23}
\end{array}\right)  
\hskip-.5em
\left(\begin{array}{ccc}
1& 0& 0\\
0&\  e^{i\phi_2}& 0\\
0& 0& e^{i(\phi_3+\delta)}
\end{array}\right) && \quad
\end{eqnarray}
where $c_{ij}=\cos\theta_{ij}$ and $s_{ij} = \sin\theta_{ij}$. The extra diagonal phases are present for Majorana neutrinos but do not affect oscillation phenomena.

With three neutrinos there are two independent $\delta m^2$ and $\delta m_a^2 \gg \delta m_b^2$ is indicated by the oscillation evidence. The vacuum oscillation probabilities are
\begin{equation}
P(\nu_\alpha \to \nu_\beta) = A_{\alpha\beta} \sin^2 \Delta_a - B_{\alpha\beta} 
\sin^2\Delta_b + \epsilon_{\alpha\beta} J \sin2\Delta_b \,,
\end{equation}
where $\Delta_a \equiv \delta m_a^2 L / 4E_\nu$. $A_{\alpha\beta}$ is the amplitude of the leading oscillation, $B_{\alpha\beta}$ the amplitude of the sub-leading oscillation and $J$ the CP-violating amplitude; all are determined by the $U$ matrix elements. The physical variable is $L/E_\nu$, where $L$ is the baseline from source to detector and $E_\nu$ is the neutrino energy.


In matter, $\nu_e$ scatter differently from $\nu_\mu$ and $\nu_\tau$, and the effective neutrino mixing amplitude in matter can be very different from the vacuum amplitude\cite{wolf}. For the leading oscillation  the matter and vacuum oscillation mixings are related in the approximation of constant matter density by
\begin{equation}
\sin^22\theta_{13}^m = {\sin^22\theta_{13}\over 
\left(\cos2\theta_{13} - A/\delta m_a^2\right)^2 + \sin^22\theta_{13}}\,,
\end{equation}
where
\begin{equation}
A = 2\sqrt2\, G_F \, Y_e\, \rho(x)\,E_\nu \,.
\end{equation}
Here $Y_e$ is the electron fraction and $\rho(x)$ is the density at path-length $x$. The $\nu_e\to\nu_\mu$ (or $\nu_e\to\nu_\tau$) oscillation argument in matter of constant density is
\begin{equation}
\Delta_a^m = {1.27\delta m_a^2({\rm eV^2})\, L({\rm km})\over E_\nu(\rm GeV)} \; \sqrt{\left({A\over\delta m_a^2} - \cos2\theta_{13}\right)^2 + \sin^22\theta_{13}}\,.
\end{equation}

Resonance enhancements in matter are possible for $\delta m_a^2 > 0$, while suppression occurs for $\delta m_a^2 < 0$. It is significant that the resonant energies correspond to neutrino energies relevant to the atmospheric and solar anomalies.
\begin{eqnarray}
&{\rm Earth:}&\quad E_\nu\simeq 15{\rm~GeV} \left(\delta m_a^2\over 3.5\times10^{-3}{\rm\,eV^2}\right) \left(1.5{\rm~gm/cm^3}\over \rho Y_e\right)
\\
&\rm Sun:&\quad E_\nu\simeq 10{\rm\ MeV} \left( \delta m_b^2\over 10^{-5}{\rm \,eV^2} \right) \left( 10{\rm~g/cm^3}\over \rho Y_e \right)\,.
\end{eqnarray}
%

\subsection{Atmospheric Neutrino Oscillations}

The Kamiokande, SuperKamiokande (SuperK), Macro and Soudan atmospheric neutrino measurements\cite{kajita} show a $\mu/e$ ratio that is about 0.6 of expectations. The SuperK experiment has established the dependence of the $e$ and $\mu$ event rates on zenith angle, or equivalently the baseline $L$. $N(e)$ is independent of $L$ and validates the $\nu_e$ flux calculation (within 20\%). $N(\mu)$ depletion increases with $L$. The muon event distributions are will described by $\nu_\mu\to\nu_\tau$ vacuum oscillations with a $\nu_\mu$ {\em survival} probability
\begin{equation}
P(\nu_\mu\to\nu_\mu) = 1 - A_{\mu\tau} \sin^2 (1.27\delta m_a^2 L/E_\nu)
\end{equation}
with $\delta m_a^2 = 3.5\times 10^{-3}\rm\,eV^2$ and maximal or near maximal amplitude
\begin{equation}
A_{\mu\tau} = 1^{+0.0}_{-0.2}\quad (i.e.,\ |\theta_{32} - 45^\circ| < 13^\circ)
\,.
\end{equation}
With further data accumulation\cite{kajita} a slightly lower central value is now indicated ($\delta m_a^2 = 2.8 \times 10^{-3}\rm\,eV^2$). The $\nu_\mu\to\nu_\tau$ oscillations are not resolved due to smearing of $L$ and inferred $E_\nu$ values, and equally good fits to the SuperK data are found with oscillation and neutrino decay ($\nu_2\to\bar\nu_4 + J$) models\cite{nu-decay}. For now we assume the simplest interpretation of the SuperK data, namely oscillations. We note that in the CHOOZ reactor experiment $\bar\nu_e$ disappearance is not observed at the $\delta m_a^2$ scale and the corresponding constraint on 3-neutrino mixing for $\delta m_a^2 = 3.5\times10^{-3}\rm\,eV^2$ is
\begin{equation}
A_{\mu e} < 0.2\,, \quad |U_{e3}| < 0.23\,, \quad \theta_{13} < 13^\circ \,.
\end{equation}
%

\subsection{Solar Neutrino Oscillations}

The solar neutrino experiments\cite{kajita} sample different $\nu_e$ energy ranges and find different flux deficits compared to the Standard Solar Model (SSM)\cite{bahcall-sm} as follows:
\begin{equation}
\def\arraystretch{1.5}
\arraycolsep=.5em
\begin{array}{lll}
\nu_e {}^{71}{\rm Ga} \to {}^{71}{\rm Ge}\,e& \begin{array}{l}\rm GALLEX\\[-1ex] \rm SAGE\end{array} & \begin{array}{l}0.60\pm 0.06\\[-1ex] 0.52\pm 0.06\end{array}\\
\nu_e {}^{37}{\rm Cl}\to {}^{37}{\rm Ar}\, e&\rm\ Homestake& \ 0.33\pm 0.03\\
\nu e\to \nu e& \rm\ SuperK& \ 0.47\pm 0.02 \end{array}
\end{equation}
Thus the $\nu_e$ survival probability is inferred to be energy dependent.

Global oscillation fits\cite{solardata} have been made using floating $^8$B and hep flux normalizations which are somewhat uncertain in the SSM. The relative normalizations from the fits range from 0.5 to 1.2 for the $^8$B flux and 1 to 25 for the hep flux. These global fits include the data on
(i)~total rates, assuming all the experiments are okay --- the different Cl suppression ratio plays a vital role;
(ii)~the night-day asymmetry, which is observed at the 2$\sigma$ level\cite{kajita} --- the large angle matter solution gives night rates $>$ day rates;
(iii)~seasonal dependence beyond $1/r^2$, which can occur for vacuum solutions.
The oscillation analyses\cite{solardata} generally agree on the allowed $\delta m_{21}^2$ and $\sin^22\theta_{12}$ regions for acceptable solutions.
Typical candidate solar solutions are given in Table~\ref{tab:solar}. In the case of vacuum oscillations (VO) several discrete regions of $\delta m_{21}^2$ are possible.

\begin{table}[h]
\vskip-4ex

\def\arraystretch{1.2}
\caption{\label{tab:solar}
Representative solutions to the solar neutrino anomaly.}
\centering\leavevmode
\begin{tabular}{lcc}
solution& $\sin^22\theta_{12}$& $\delta m^2_{21}\rm\ (eV^2)$\\
\hline
SAM& $\sim5\times10^{-3}$& $\sim5\times10^{-6}$\\
LAM& $\sim1$& $\sim3\times10^{-5}$\\
LOW& $\sim1$& $\sim10^{-7}$\\
VO& $\sim1$& $\sim10^{-10}$\\
\hline
\end{tabular}
\end{table}

\subsection{3-Neutrino Mixing Matrix}

Once the solar oscillation solution is pinned down, and $\theta_{12}$ is thus determined, we will have approximate knowledge of the mixing angles of the 3-neutrino matrix, with $\theta_{23} \sim \pi/4$ and $\theta_{13} \sim 0$ from the atmospheric and CHOOZ data. Upcoming experiments are expected to shed light on the solar solution. In the SNO experiment, which is now taking data, and the forthcoming ICARUS experiment, the high energy $\nu_e$ CC events may distinguish LAM, SAM, and LOW solutions with large hep flux contributions from the VO or the SAM sterile neutrino solutions\cite{BKS99}. Also, the neutral-current to charged-current ratio will distinguish active from sterile oscillations. The Borexino experiment can measure the VO seasonal variation of the $^7$Be line flux. The KamLand reactor experiment to measure the $\bar\nu_e$ survival probability will be sensitive to the LAM and LOW solar solutions\cite{muray-kam}.

The CP phase $\delta$ may be measurable at a neutrino factory if the solar solution is LAM\cite{euro-cp,BGRW}. The CP violation comes in only at the sub-leading oscillation scale\cite{BWP-P}.
An apparent CP-odd asymmetry is induced by matter.

\subsection{Models}

For maximal mixing in both atmospheric and solar sectors, there is an unique mixing matrix\cite{bimax1} 
\begin{equation}
U = 
\left(\begin{array}{ccc}
1/\sqrt 2& -1/\sqrt 2& 0\\ 1/2& 1/2& -1/\sqrt 2\\ 1/2& 1/2& 1/\sqrt2
\end{array}\right) \,.
\end{equation}
In this bimaximal mixing model, there would be no CP-violating effects. However, because $U_{e3} = 0$, long-baseline experiments would have some sensitivity to the sub-leading LAM solar scale oscillations.

Many unification models predict that the neutrino masses are Majorana and hierarchical, there is no cosmologically significant dark matter, and the SAM solar solution (small $\theta_{12}$ mixing) obtains\cite{RMreview,king}. 

\subsection{Beyond 3 Neutrinos}

The LSND evidence for $\nu_\mu\to\nu_e$ oscillations with $\delta m^2 \sim 1\rm~eV^2$, $\sin^22\theta\sim 10^{-2}$ requires a $\delta m^2$ scale distinct from the atmospheric and solar oscillation scales, and thus a sterile neutrino state would be needed to explain all the oscillation phenomena. Then to also satisfy limits from CDHS accelerator and Bugey reactor experiments, the mass hierarchy must be two separated pairs\cite{doublets}. Such a scenario would allow even more interesting effects at a neutrino factory, such as large CP violation, since both the leading and sub-leading oscillation scales would be accessible. The MiniBooNE experiment will settle whether the LSND evidence is real. Other interest in sterile neutrinos comes from $r$-process nucleosynthesis if it occurs in supernovae\cite{baha}.

\section{Long-Baseline Experiments}

Long-baseline experiments are needed to 
(i)~confirm the atmospheric evidence for $P(\nu_\mu\to\nu_\mu)$ at accelerators;
(ii)~resolve the leading $\nu_\mu\to\nu_\mu$ oscillation and exclude the neutrino decay possibility;
(iii)~precisely measure $|\delta m_a^2|$;
(iv)~exclude $\nu_\mu\to\nu_s$ disappearance, although SuperK has now shown that oscillations to sterile neutrinos are excluded at 99\% CL\cite{kajita};
(v)~measure $|U_{e3}|$ from $\nu_e\to\nu_\mu$ appearance, which requires a muon decay source for the neutrino beam; 
(vi)~determine the sign of $\delta m_a^2$ from matter effects in the Earth's crust; and
(vii)~search for CP violation.


The first long-baseline experiments will measure the energy dependence of the produced muons and measure the neutral-current to charged-current ratio, to partially address the first four issues listed above. The K2K experiment from KEK to SuperK is in operation, with a baseline $L=250$~km and mean neutrino energy $\left< E_\nu \right> = 1.4$~GeV. The MINOS experiment from Fermilab to Soudan, with $L=732$~km and possible energies of $\left<E_\nu\right> = 3, 6, 12$~GeV will begin in 2002. A 10\% precision on $|\delta m_a^2|$ may ultimately be possible at MINOS. The ICANOE and OPERA long-baseline experiments from CERN to Gran Sasso with $L\simeq 743$~km have been approved.


Muon storage rings could provide intense neutrino beams ($\sim 10^{19}\mbox{--}10^{21}$ per year) that would yield thousands of charged-current neutrino interactions in a reasonably sized detector (10--50~kt) anywhere on Earth\cite{nufactories,nufactories2}. These neutrino factories would have pure neutrino beams ($\nu_e,\bar\nu_\mu$ from stored $\mu^+$ and $\bar\nu_e, \nu_\mu$ from stored $\mu^-$) with 50\% $\nu_e$ or $\bar\nu_e$ components. Detection of wrong-sign muons (the muons with opposite sign to the charge current from the beam muon neutrino) would signal $\nu_e\to\nu_\mu$ or $\bar\nu_e\to\bar\nu_\mu$ appearance oscillations. We now discuss the capability of a neutrino factory with $2\times10^{20}$ muons a year and a 10~kt detector to resolve the issues raised in the preceding section.

With  an $E_\mu=30$~GeV storage ring at a baseline of $L=2800$~km, a statistical precision of a few \% on $\sin^22\theta_{23}$ is possible in $\nu_\mu$ survival measurements. This accuracy in measuring $\sin^22\theta_{23}$ would differentiate the bimaximal model prediction\cite{bimax1} of $\sin^22\theta_{23}=1$ from the democratic model prediction\cite{democ} of $\sin^22\theta_{23} = 8/9$.

With stored muon energies $E_\mu = 10$ to 50~GeV and baselines of $L = 732$ to  7332~km and 1~kt detector, there would be hundreds of events per year from $\nu_\mu\to\nu_\tau$ oscillations  for an intensity of $2\times10^{20}$ neutrinos.

The wrong-sign muon event rates are approximately proportional to $\sin^22\theta_{13}$. 
For non-zero $\sin^22\theta$ the observation of $\nu_e\to\nu_\mu$ and $\bar\nu_e\to\bar\nu_\mu$ appearance oscillations at baselines long enough to have significant matter effects will allow\cite{BGRW,nufactories2,shrock} a determination of the sign of $\delta m_{32}^2$, and thus determine the pattern of the masses (a $1+2$ mass hierarchy versus a $2+1$ hierarchy for three neutrinos). A proof of the principle that the sign of $\delta m^2$ can be so determined has been given\cite{BGRW,nufactories2} for a baseline $L = 2800$~km. In $\mu^+$ appearance, $\delta m_{32}^2 >0$ gives a smaller rate and harder spectrum than $\delta m_{32}^2 < 0$, while the results are opposite in $\mu^-$ appearance\cite{nufactories2}.

In optimizing $E_\mu$ and $L$ for long-baseline experiments to find the sign of $\delta m_{32}^2$, $L = 732$~km is too short (matter effects are small) and $L = 7332$~km is too far (event rates are low). The sensitivity to determine the sign of $\delta m_{32}^2$ improves linearly with $E_\mu$. There is a tradeoff between energy, detector size and muon beam intensity\cite{BGRW}.

\section{Absolute Neutrino Masses}

Oscillation phenomena determine only mass-squared differences, leaving the absolute mass scale unknown. However, because the atmospheric and solar $\delta m^2$ values are $\ll (1\rm~eV)^2$, all mass eigenvalues are approximately degenerate if at the $\sim 1$~eV scale. Thus all neutrino mass eigenvalues are bounded\cite{BWW} by the tritium limit from the Troitsk and Mainz experiments,
\begin{equation}
m_j < 3{\rm\ eV\quad for}\ j = 1,2,3 \,.
\end{equation}

Neutrinoless double-beta decay (0$\nu\beta\beta$) provides a probe of Majorana neutrino mass\cite{0nbb}. The rate is proportional to the $\nu_e\nu_e$ element of the neutrino mass matrix. The present limit from the Heidelberg experiment\cite{baudis} is 
\begin{equation}
M_{\nu_e\nu_e} < 0.2{\rm\ eV} \times f\,,
\end{equation}
where the factor $f$ represents uncertainty in the nuclear matrix elements, which might be as large as a factor of~3. The $0\nu\beta\beta$ limit translates to a bound on the summed neutrino Majorana masses of\cite{BW99}
\begin{equation}
\sum m_\nu < 0.75{\rm\ eV}\times f
\end{equation}
in the SAM solar solution. No similar constraints apply to the LAM, LOW or VO solutions where the bound can be satisfied by having opposite CP parity of $\nu_1$ and $\nu_2$ mass eigenstates. Future experiments\cite{future} may probe to
\begin{equation}
|M_{\nu_e\nu_e}| = 0.01\rm\ eV \,,
\end{equation}
 which would provide sensitivity down to
\begin{equation}
\sum m_\nu = 0.08{\rm\ eV} \times f
\end{equation}
in the SAM solution.


Measurements of the power spectrum by the MAP and PLANCK satellites may determine $\sum m_\nu$ down to $\sim 0.4$~eV~\cite{cosmo}. The heights of the acoustic peaks can also decide how the mass is distributed among the neutrino eigenstates\cite{falk}.

\section{Summary}

We have entered an exciting new era in the study of neutrino masses and mixing. From the SuperK evidence on atmospheric neutrino oscillations, we already have a surprising amount of information about the neutrino mixing matrix (near maximal $\sin^22\theta_{23}$ and near minimal $\sin^22\theta_{13}$). The SuperK, SNO, Borexino, KamLand, and ICARUS experiments are expected to differentiate among the candidate solar oscillation possibilities and determine $\sin^22\theta_{12}$. MiniBooNE will tell us whether a sterile neutrino is mandated. Neutrino factories will study the leading oscillations, determine the sign of $\delta m_a^2$, measure $U_{e3}$, and possibly detect CP violation. The GENIUS $0\nu\beta\beta$ experiment and the MAP and PLANCK satellite measurements of the power spectrum will probe the absolute scale of neutrino masses. There is a synergy of particle, physics, nuclear physics, and cosmology occurring in establishing the fundamental properties of neutrinos. A theoretical synthesis should emerge from these experimental pillars.

A more complete version of this review, including figures and more extensive references, can be found in Ref.~\citenum{mumu99}.

\section*{Acknowledgments}

I thank my collaborators K.~Whisnant, T.~Weiler, S.~Pakvasa, S.~Geer, R.~Raja, J.~Learned, P.~Lipari, and M.~Lusignoli. This research was supported in part by the U.S. Department of Energy under
Grant  No.~DE-FG02-95ER40896  and in part by the
University of Wisconsin Research Committee with funds granted by the Wisconsin
 Alumni Research Foundation.

\end{document}